\documentclass[%
 reprint,
nofootinbib,
 amsmath,amssymb,
 aps,
]{revtex4-1}

\usepackage[dvipdfmx]{graphicx}
\usepackage[dvipdfmx]{}
\usepackage{dcolumn}
\usepackage{bm}

\usepackage{multirow}

\usepackage{feynmf}
\usepackage{here}
\usepackage{comment}

\usepackage{ulem}
\usepackage[usenames]{color}
\usepackage{float}
\usepackage{color}

\newcommand{\Slash}[1]{{\ooalign{\hfil/\hfil\crcr$#1$}}}

\begin{document}

\title{Analysis of $\Lambda_c(2595)$, $\Lambda_c(2625)$, $\Lambda_b(5912)$, $\Lambda_b(5920)$ based on a 
chiral partner structure}

\author{Yohei Kawakami} \email{kawakami@hken.phys.nagoya-u.ac.jp}
 \affiliation{Department of Physics, Nagoya University, Nagoya, 464-8602, Jpana}
 \author{Masayasu Harada}\email{harada@hken.phys.nagoya-u.ac.jp}
 \affiliation{Department of Physics, Nagoya University, Nagoya, 464-8602, Jpana}

\date{\today}

\begin{abstract}
We construct an effective hadronic model including 
$\Lambda_c(2595)$, $\Lambda_c(2625)$, $\Lambda_b(5912)$ and $\Lambda_b(5920)$
regarding them as chiral partners to $\Sigma_c(2455)$, $\Sigma_c(2520)$, $\Sigma_b$ and $\Sigma_b^\ast$, respectively,
with respecting the chiral symmetry and heavy-quark spin-flavor symmetry.
We determine the model parameters from the experimental data for relevant masses and 
decay widths of $\Sigma_c^{(\ast)}$ and $\Lambda_c(2595)$.
Then, we study the decay widths of $\Lambda_c(2625)$, $\Lambda_b(5912)$ and $\Lambda_b(5920)$.
We find that, although the decay of $\Lambda_c(2595)$ is dominated by the resonant contribution through $\Sigma_c(2455)$, non-resonant contributions are important for $\Lambda_c(2625)$, $\Lambda_b(5912)$ and $\Lambda_b(5920)$, which reflects the chiral partner structure.
We also study the radiative decays of the baryons, and show that each of their widths is determined from the  radiative decay width of their chiral partners.
\end{abstract}

\maketitle

\section{\label{sec:level1}Introduction}

Chiral symmetry and its spontaneous breaking is one of the most important properties to understand the structures of hadrons including light quarks. The spontaneous chiral symmetry breaking is expected to generate a part of hadron masses and causes mass difference between chiral partners.
We expect that the study of chiral partner structure will provide a clue for understanding the chiral symmetry.

In Refs.~\cite{Nowak:1992um,Nowak:1993vc,Bardeen:1993ae,Bardeen:2003kt,Nowak:2003ra}, the chiral partner structure of heavy-light mesons was studied regarding the mesons with $J^P = (0^+,1^+)$ such as $(D_0^\ast, D_1)$ as the chiral partners to the mesons with $J^P = (0^-,1^-)$ such as $ (D,D^\ast)$ based on the chiral symmetry combined with the heavy quark spin symmetry.
In Refs.~\cite{Ma:2015lba,Ma:2015cfa,Ma:2017nik},
doubly heavy baryons with negative parity were studied by regarding them as chiral partners to the positive parity heavy baryons.
In these analysis, the heavy quark flavor symmetry  
in addition to the chiral symmetry and the heavy quark spin symmetry plays a very important role to relate the charm baryons to the bottom baryons.
In Refs.~\cite{Nowak:2004jg,Harada:2012dm}, chiral partner structure of heavy baryons including a charm quark is within the bound 
state  approach based on the Skyrm model. 
In Ref.~\cite{Liu:2011xc}, the chiral partner structure of single heavy baryons was studied, in which the chiral partner of $\Sigma_c$ baryon with positive parity is regarded as the $\Sigma_c$ baryons with negative parity.  

In the present work, 
we would like to propose a new possibility of the chiral partner structure for single heavy baryons differently from the one  in Ref.~\cite{Liu:2011xc}, in which the chiral partners of $\Sigma_{Q}$ ($Q=c,b$) baryons with positive parity are considered as $\Lambda_Q$ baryons with negative parity:
we ragard $\left( \Lambda_c(2595;J^P = 1/2^-) \,,\, \Lambda_c(2625;3/2^-) \right)$ as the chiral partners to 
$\left( \Sigma_c(2455;1/2^+) \,,\, \Sigma_c(2520;3/2^+) \right)$, and $\left( \Lambda_b(5912;1/2^-) \,,\, \Lambda_b(5920;3/2^-) \right)$ to $\left( \Sigma_b(1/2^+) \,,\, \Sigma_b^\ast(3/2^+) \right)$.
Based on this chiral partner structure, we construct an effective model respecting the chiral symmetry and the heavy-quark spin-flavor symmetry.
Determining model parameters 
from the experimental data for relevant masses and 
decay widths of $\Sigma_c(2455)$, $\Sigma_c^{\ast}(2520)$ and $\Lambda_c(2595)$,
we study the decay widths of $\Lambda_c^\ast(2625)$, $\Lambda_b(5912)$ and $\Lambda_b(5920)$.

This paper is organized as follows:
In section~\ref{sec:chiral}, we study the chiral structure of single heavy baryons (SHBs).
We construct an effective Lagrangian in section~\ref{sec:Lag}.
Sections~\ref{sec:masses} and \ref{sec:2pi} are 
devoted to study the masses and
the hadronic decays of SHBs.
We also study the radiative decays of SHBs in section~\ref{sec:rad}.
Finally, we give a summary and discussions in section~\ref{sec:summary}.

\section{Chiral structure of single heavy baryons}
\label{sec:chiral}

In this section, we study the chiral structure of single heavy baryons using interpolating quark fields.

First we consider interpolating field operators made from up or down quarks:
\begin{equation}
q_{L,R}^{i} \ , \quad ( i = u,\, d ) \ ,
\end{equation}
where $L$ and $R$ denote left-handed and right-handed chirality, respectively.
By using these, we can construct two combinations of diquarks carrying spin-zero which are expressed as
\begin{equation}
\left( q_L^i \right)^{T} C \, q_L^j \ , \quad
\left( q_R^i \right)^{T} C \, q_R^j \ ,
\end{equation}
where $^T$ denote the transposition in the spinor space and $C = i \gamma^0 \gamma^2$ is the charge conjugation matrix.
When the relative angular momentum between two quarks are even, the indices $i$ and $j$ should be anti-symmetrized due to the Fermi statistics.
In such a case, we can easily see that both of the above diquarks are chiral singlet.
For clarifying this chiral structure, we introduce the following two diquarks which are chiral singlet:
\begin{equation}
\epsilon_{ij} \left( q_L^i \right)^{T} C\, q_L^j \ , \quad
\epsilon_{ij}  \left( q_R^i \right)^{T} C\, q_R^j \ ,
\end{equation}
where $\epsilon^{ij}$ is anti-symmetric tensor, $\epsilon_{ij} = - \epsilon_{ji}$, with $\epsilon_{ud} = 1$, and the summations over repeated indices are understood.
Since both of the above two diquarks are chiral singlet, two combination of them, which are parity eigenstates, are separately chiral singlet.

Now, let us introduce a field  
for the chiral-singlet light-quark cloud with $J^P = 0^+$ as
\begin{equation}
\Phi_{(+)} =  \epsilon_{ij} \left( q_L^i \right)^{T} C\, q_L^j + \epsilon_{ij}  \left( q_R^i \right)^{T} C\, q_R^j \ ,
\end{equation}
which belongs to $(\boldmath{1}\,,\,\boldmath{1})$ representation under $(\mbox{SU}(2)_L\,,\,\mbox{SU}(2)_R)$ symmetry.
We construct a single heavy baryon by combining this light-quark cloud ($J^P = 0^+$) to a heavy quark $Q$ ($Q = c, b$).  The resultant baryon is a heavy-quark spin singlet, so that we identify it with the lightest $\Lambda_Q$ ($Q=c,b$):
\begin{equation}
\Lambda_Q \sim Q \, \Phi_{(+)} \ , \quad 
\Lambda_Q = \left( \Lambda_c^+ \,,\, \Lambda_b^0 \right) \ ,
\label{Lambda field}
\end{equation}
which belongs to $(\boldmath{1}\,,\,\boldmath{1})$ representation under $(\mbox{SU}(2)_L\,,\,\mbox{SU}(2)_R)$ symmetry.

Next, we consider the following diquark:
\begin{equation}
\left[ \Phi^\mu \right]^{ij} =
	\left[ q_{L}^{T}C\gamma^{\mu}q_{R} \right]^{ij} = 
	\left( q_{L}^i \right)^{T}C\gamma^{\mu}q_{R}^j  \ ,
\end{equation}
which belongs to $(\boldmath{2}\,,\,\boldmath{2})$ representation under $(\mbox{SU}(2)_L\,,\,\mbox{SU}(2)_R)$ symmetry.
We can easily see that the following property is satisfied:
\begin{equation}
\left[ q_{R}^{T}C\gamma^{\mu}q_{L} \right]^{ij} = - \left[ q_{L}^{T}C\gamma^{\mu}q_{R}  \right]^{ji} \ .
\label{transpose}
\end{equation}
From these diquarks, we make two combinations of parity eigenstates:
\begin{align}
\left[ q_L^T C \gamma^\mu q_R \right]^{ij} + \left[ q_R^T C \gamma^\mu q_L \right]^{ij}  & = \left[ q^T C \gamma^\mu q \right]^{ij}
= \left[ \Phi_{(3)}^\mu \right]^{ij} \ , \notag \\
\left[ q_L^T C \gamma^\mu q_R \right]^{ij} - \left[ q_R^T C \gamma^\mu q_L \right]^{ij}  & = \left[ q^T C \gamma^\mu \gamma_5 q \right]^{ij}
= \left[ \Phi_{(1)}^\mu \right]^{ij} \ .
\end{align} 
From the property in Eq.~(\ref{transpose}), 
one can easily check that the indices of 
the diquark with $J^P = 1^+$ is symmetric in the light-quark flavor space,  and those of the one with $J^P = 1^-$ is anti-symmetric, i.e. 
\begin{align}
\left[ \Phi_{(3)}^\mu \right]^{ij} = & \left[ \Phi_{(3)}^\mu \right]^{ji} 
\ , \notag \\
\left[ \Phi_{(1)}^\mu \right]^{ij} = & - \left[ \Phi_{(1)}^\mu \right]^{ji} 
 \ . 
\end{align}
From this we can easily see that, 
when the chiral symmetry is spontaneously broken into the isospin symmetry, 
$\Phi_{(3)}^\mu$
is the iso-triplet diquark with $J^P = 1^+$, and 
$\Phi_{(1)}^\mu$
is the iso-singlet diquark with $J^P = 1^-$. 

The diquark $\Phi^\mu$ combined with a heavy quark makes a set of heavy-quark doublets of single heavy baryons (SHBs) with $1/2^-$ and $3/2^-$ as
\begin{equation}
S_Q^\mu \sim Q \, \Phi^\mu \ ,
\label{chiral fields}
\end{equation}
where $S_Q^\mu$ denotes the field for the set of SHBs.
The $S_Q^\mu$ includes 
iso-triplet SHBs and iso-singlet SHBs as 
\begin{align}
& \left( \Sigma_{Q}^a(1/2^+) \,,\, \Sigma_{Q}^{\ast a} (3/2^+) \right) \sim Q \, \Phi^\mu_{(3)} \ , \notag\\
& \left( \Lambda_{Q1}(1/2^-) \,,\, \Lambda_{Q1}^{\ast}(3/2^-) \right) \sim Q  \, \Phi^\mu_{(1)} \ ,
\end{align}
where we omitted the index $\mu$ in the left hand sides.
It should be stressed that, since both $\Phi_{(3)}^\mu$ and $\Phi_{(1)}^\mu$ are included in one chiral multiplet $\Phi^\mu$, the heavy quark multiplet of $\left( \Lambda_{Q1}(1/2^-) \,,\, \Lambda_{Q1}^{\ast}(3/2^-) \right)$  is the chiral partner to that of $\left( \Sigma_{Q}(1/2^+) \,,\, \Sigma_{Q}^{\ast} (3/2^+) \right)$.
In the present work, we identify $\left( \Sigma_{Q}(1/2^+) \,,\, \Sigma_{Q}^{\ast} (3/2^+) \right)$ with the lightest iso-triplet single-heavy baryons with positive parity, and $\left( \Lambda_{Q1}(1/2^-) \,,\, \Lambda_{Q1}^{\ast}(3/2^-) \right)$ with the lightest iso-singlet ones with negative parity:
\begin{align}
\left( \Sigma_{c}\,,\, \Sigma_{c}^{\ast} \right) = &\left( \Sigma_c(2455;1/2^+) \,,\, \Sigma_c(2520;3/2^+) \right) \ , \notag \\
\left( \Lambda_{c1} \,,\, \Lambda_{c1}^{\ast} \right) = & \left( \Lambda_c(2595;1/2^-) \,,\, \Lambda_c(2625;3/2^-) \right) \ , \notag\\
\left( \Sigma_{b}\,,\, \Sigma_{b}^{\ast} \right) = & \left( \Sigma_b(1/2^+) \,,\, \Sigma_b^\ast(3/2^+) \right) \ , \notag\\
\left( \Lambda_{b1} \,,\, \Lambda_{b1}^{\ast} \right) = & \left( \Lambda_b(5912;1/2^-) \,,\, \Lambda_b(5920;3/2^-) \right)\ .
\end{align}

\section{Effective Lagrangian}
\label{sec:Lag}

In this section we construct an effective Lagrangian for the relevant single heavy baryons (SHBs) based on the heavy-quark spin-flavor symmetry and the chiral symmetry.
We use the field $\Lambda_Q$ for expressing the SHBs belonging to the chiral singlet 
in Eq.~(\ref{Lambda field}).
For expressing the SHBs belonging to chiral $ (2, 2) $ representations we introduce the field $S_Q^\mu$ in Eq.~(\ref{chiral fields}) which transforms as
\begin{equation}
	S^{\mu}_{Q}\stackrel{\mathrm{Ch.}}{\to} g_{R}S^{\mu}_{Q}g_{L}^{T},\quad(Q=c,\ b).
\end{equation}
As we discussed in the previous section, 
we assume that the fields include the iso-triplet SHBs with positive parity and the iso-singlet SHBs with negative parity as chiral partners to each others. They are embedded into the field $S^\mu_Q$ as
\begin{equation}
	S^{\mu}_{Q}=\hat{\Sigma}_{Q}^{\mu}+\hat{\Lambda}_{Q1}^{\mu}\ ,
\end{equation}
where $\hat{\Sigma}^\mu_Q$ and $\hat{\Lambda}^\mu_{Q1}$ 
include the iso-triplet and iso-singlet fields, respectively as
\begin{equation}
\hat{\Sigma}^{\mu}_{Q}
=\begin{pmatrix}\Sigma^{I=1\mu}_{Q} & \frac{1}{\sqrt{2}}\Sigma^{I=0\mu}_{Q}\\
	\frac{1}{\sqrt{2}}\Sigma^{I=0\mu}_{Q} & \Sigma^{I=-1\mu}_{Q} \end{pmatrix},
\end{equation}
\begin{equation}
\hat{\Lambda}^{\mu}_{Q1}
=\begin{pmatrix}0 & \frac{1}{\sqrt{2}}\Lambda^{\mu}_{Q1}\\
-\frac{1}{\sqrt{2}}\Lambda^{\mu}_{Q1} & 0 \end{pmatrix}\ .
\end{equation}
These 
$\Sigma_{Q}^{\mu}$ and $\Lambda^{\mu}_{Q1}$ are 
decomposed into spin-$3/2$ baryon fields and spin-$1/2$ fields as 
\begin{align}
	\Sigma^{\mu}_{Q}=& \Sigma^{*\mu}_{Q}-\frac{1}{\sqrt{3}}(\gamma^{\mu}+v^{\mu})\gamma_{5}\Sigma_{Q}\ , \\
		\Lambda^{\mu}_{Q1}=& \Lambda^{*\mu}_{Q1}-\frac{1}{\sqrt{3}}(\gamma^{\mu}+v^{\mu})\gamma_{5}\Lambda_{Q1}\ , 
\end{align}
where $\Sigma^{*\mu}_{Q}$ and $\Lambda^{* \mu}_{Q1}$ denote the spin-$3/2$ baryon fields, and $\Sigma_Q$ and $\Lambda_{Q1}$ the spin-$1/2$ fields, respectively.
We note that the parity transformation of the $S^\mu_Q$ field is given by
\begin{equation}
	S^{\mu}_{Q}\stackrel{\mathrm{P}}{\to}-\gamma^{0}S_{Q\mu}^{T}\ ,
\end{equation}
where $^T$ denotes the transposition of the $2\times 2$ matrix in the light-quark flavor space, and that the Dirac conjugate is defined as
\begin{equation}
	\bar{S}^{\mu}_{Q}=S^{\mu\dagger}_{Q}\gamma^{0}.
\end{equation}

We introduce a $2\times2$ matrix field $M$ for scalar and pseudoscalar mesons including a light quark and a light anti-quark, which belongs to the $(2,2)$ representation under the chiral SU$(2)_L \times$SU$(2)_R$ symmetry.
The transformation properties of $M$ under the chiral symmetry and the parity are given by
\begin{align}
	M& \stackrel{\mathrm{Ch.}}{\to}g_{L}Mg_{R}^{\dagger}\ , \\
	M& \stackrel{\mathrm{P}}{\to}M^{\dagger}\ .
\end{align}
We assume that the effective Lagrangian terms for $M$ are constructed in such a way that the $M$ has a vacuum expectation value (VEV) which breaks the chiral symmetry spontaneously, and the VEV is proportional
to the pion decay constant $f_\pi$~\footnote{Here we adopt the normalization of $f_\pi = 92.4\,$MeV.
}:
\begin{equation}
\langle M \rangle = f_\pi \begin{pmatrix} 1 & 0 \\ 0 & 1 \end{pmatrix} \ .\label{vev}
\end{equation}
In the following, for studying the decays of the single heavy baryons with emitting pions, we parameterize the field $M$ as
\begin{equation}
M = f_\pi \, U \ ,
\end{equation}
where
\begin{equation}
	U=e^{\frac{2 i\pi}{f_{\pi}}},
\end{equation}
with $\pi$ being the $2\times2$ matrix field including pions as
\begin{equation}
\pi=\frac{1}{2} \, \begin{pmatrix} \pi^{0} & \sqrt{2}\pi^{+}\\
		\sqrt{2}\pi^{-} & -\pi^{0} \end{pmatrix}.
\end{equation}

\begin{widetext}
Now, let us write down an effective Lagrangian including the baryon fields $\Lambda_Q$ and $S_{Q}^{\mu}$ together with the meson field $M$, based on the heavy-quark spin-flavror symmetry and the chiral symmetry.
We do not include the terms including more than square of $M$ field or more than two derivatives.
A possible Lagrangian is given by
\begin{align}
	\mathcal{L}_{Q}\notag=&-{\rm{tr}}\bar{S}_{Q}^{\mu}\left(v\cdot iD-\Delta_{Q}\right)S_{Q\mu}+
\bar{\Lambda}_{Q}\left(v\cdot iD\right)\Lambda_{Q}
\\\notag
	&+\frac{g_{1}}{2f_{\pi}}{\rm{tr}}\left(\bar{S}_{Q}^{\mu}M^{\dagger}MS_{Q\mu}+\bar{S}_{Q\mu}^{T}MM^{\dagger}S_{Q}^{\mu T}\right)\\ \notag
	&-\frac{g_{2}}{2f_{\pi}}{\rm{tr}}\bar{S}_{Q}^{\mu}M^{\dagger}S_{Q\mu}^{T}M^{T}
	-\frac{g_{2}^{v}}{2m_{\Lambda_{Q}}}{\rm{tr}}\bar{S}_{Q}^{\mu}M^{\dagger}S_{Q\mu}^{T}M^{T}\\\notag
	&-i\frac{h_{1}^{I}-ih_{1}^{R}}{4f_{\pi}^{2}}{\rm{tr}}\left(\bar{S}_{Q}^{\mu}M^{\dagger}v\cdot\partial MS_{Q\mu}+\bar{S}_{Q}^{\mu T}Mv\cdot\partial M^{\dagger}S_{Q\mu}^{T}\right)
	\notag\\
	&
	-i\frac{-h_{1}^{I}-ih_{1}^{R}}{4f_{\pi}^{2}}{\rm{tr}}\left(\bar{S}_{Q}^{\mu}v\cdot\partial M^{\dagger}MS_{Q\mu}+\bar{S}_{Q}^{\mu T}v\cdot\partial MM^{\dagger}S_{Q\mu}^{T}\right)\notag\\ \notag
	&+\frac{h_{2}}{2f_{\pi}^{2}}{\rm{tr}}\left(\bar{S}_{Q}^{\mu}v\cdot\partial M^{\dagger}S_{Q\mu}^{T}M^{T}+\bar{S}_{Q}^{\mu T}v\cdot\partial MS_{Q\mu}M^{*}\right)\\ 
	&-\frac{g_{3}}{2\sqrt{2}f_{\pi}}\bar{\Lambda}_{Q}{\rm{tr}}\left(\partial^{\mu}MS_{Q\mu}\tau^{2}-\partial_{\mu}M^{\dagger}S_{Q}^{\mu T}\tau^{2}\right)+{\rm{h.c.}},
\end{align}
where $m_{\Lambda_{Q}}$ ($Q=c,b$) are the masses of $\Lambda_{c}(2286)$ and $\Lambda_{b}$ in the ground state,
$\Delta_Q$ provides the difference between the chiral invariant masses of $(\Sigma_Q,\Lambda_{Q1})$ chiral multiplet and the chiral singlet $\Lambda_Q$ with heavy-quark flavor violation included. $g_i\ (i=1,2,3)$, $g_2^v$, $h_1^I$, $h_1^R$ and $h_2$ are dimensionless coupling constants.
Note that we included $g_2^v$-term to incorporate
the heavy-flavor violation needed for explaining the mass differences of charm and bottom sectors.
Although we can add heavy-quark flavor violation terms corresponding to $g_{1}$-term, such contributions are absorbed into the definition of $\Delta_Q$.
We expect that heavy-quark flavor violating corrections to other terms are small.

\end{widetext}

\section{Masses and $\Sigma_{Q}^{(*)}\to\Lambda_{Q}\pi$ decays}
\label{sec:masses}

In this section, we determine the coupling constants $g_{2}$ and $g_{2}^{v}$ from masses of relevant heavy baryons, and $g_{3}$ from $\Sigma_{c}^{(*)}\to\Lambda_{c}\pi$ decays. Then we make predictions of $\Sigma_{b}^{(*)}\to\Lambda_{b}\pi$ decays. 

When the chiral symmetry is spotaneously broken, the light meson field $M$ acquires its vacuum expectation value as in Eq.~(\ref{vev}). 
Then the masses of $\Sigma_{Q}^{(*)}$ and $\Lambda_{Q1}^{(*)}$ are expressed as 
\begin{equation}
	m(\Sigma_{Q}^{(*)})=m_{\Lambda_{Q}}+\Delta_{Q}+g_{1}f_{\pi}-\frac{g_{2}^{Q}}{2}f_{\pi},
\end{equation}
\begin{equation}
	m(\Lambda_{Q1}^{(*)})=m_{\Lambda_{Q}}+\Delta_{Q}+g_{1}f_{\pi}+\frac{g_{2}^{Q}}{2}f_{\pi},
\end{equation}
where $g_{2}^{Q}$ is 
\begin{equation}
	g_{2}^{Q}=g_{2}+g_{2}^{v}\frac{f_{\pi}}{m_{\Lambda_{Q}}}.
\end{equation}
In the present analysis, we assume that the heavy-quark multiplet of 
$\left(\Lambda_{c1},\ \Lambda_{c1}^{*}\right)=\break\left(\Lambda_{c}(2595;\ J^{P}=1/2^{-}),\ \Lambda_{c}(2625;\ 3/2^{-})\right)$
is the chiral partner to the multiplet of $\left(\Sigma_{c},\ \Sigma_{c}^{*}\right)=\break\left(\Sigma_{c}(2455;\ 1/2^{+}),\ \Sigma_{c}(2520;\ 3/2^{+})\right)$, and that
$\left(\Lambda_{b1},\ \Lambda_{b1}^{*}\right)=\break\left(\Lambda_{b}(5912;\ 1/2^{-}),\ \Lambda_{b}(5920;\ 3/2^{-})\right)$
to $\left(\Sigma_{b},\ \Sigma_{b}^{*}\right)=\break\left(\Sigma_{b}(1/2^{+}),\ \Sigma_{b}(3/2^{+})\right)$.
We list experimental data of
their masses and full decay widths \cite{PDG2016} in Table~\ref{exp. table}.
\begin{table}[H]
	\caption{Experimental data of masses and decay widths of heavy baryons included in the present analysis}
	\begin{center}
		\begin{tabular}{ccccc} \hline \hline
		 particle & $J^{P}$ & mass[MeV] & full width[MeV]\\\hline
		 $\Lambda_{c}$ & $1/2^{+}$ & $2286.46\pm0.14$ & no strong decays\\\hline
		 $\Sigma_{c}^{++}(2455)$ & $1/2^{+}$ & $2453.97\pm0.14$ & $1.89^{+0.09}_{-0.18}$\\
		 $\Sigma_{c}^{+}(2455)$ & $1/2^{+}$ & $2452.9\pm0.4$ & $<4.6$\\
		 $\Sigma_{c}^{0}(2455)$ & $1/2^{+}$ & $2453.75\pm0.14$ & $1.83^{+0.11}_{-0.19}$\\\hline
		 $\Sigma_{c}^{++}(2520)$ & $3/2^{+}$ & $2518.41^{+0.21}_{-0.19}$ & $14.78^{+0.30}_{-0.40}$\\
		 $\Sigma_{c}^{+}(2520)$ & $3/2^{+}$ & $2517.5\pm1.3$ & $<17$\\
		 $\Sigma_{c}^{0}(2520)$ & $3/2^{+}$ & $2518.48\pm0.20$ & $15.3^{+0.4}_{-0.5}$\\\hline
		 $\Lambda_{c}(2595)$ & $1/2^{-}$ & $2595.25\pm0.28$ & $2.59\pm0.30\pm0.47$\\
		 $\Lambda_{c}(2625)$ & $3/2^{-}$ & $2628.11\pm0.19$ & $<0.97$\\\hline
		 $\Lambda_{b}$ & $1/2^{+}$ & $5619.58\pm0.17$ & no strong decays\\\hline
		 $\Sigma_{b}^{+}$ & $1/2^{+}$ & $5811.3^{+0.9}_{-0.8}\pm1.7$ & $9.7^{+3.8}_{-2.8}\ ^{+1.2}_{-1.1}$\\
		 $\Sigma_{b}^{0}$ & $1/2^{+}$ & - & -\\
		 $\Sigma_{b}^{-}$ & $1/2^{+}$ & $5815.5^{+0.6}_{-0.5}\pm1.7$ & $4.9^{+3.1}_{-2.1}\pm1.1$\\\hline
		 $\Sigma_{b}^{*+}$ & $3/2^{+}$ & $5832.1\pm0.7\ ^{+1.7}_{-1.8}$ & $11.5^{+2.7}_{-2.2}\ ^{+1.0}_{-1.5}$\\
		 $\Sigma_{b}^{*0}$ & $3/2^{+}$ & - & -\\
		 $\Sigma_{b}^{*-}$ & $3/2^{+}$ & $5835.1\pm0.6\ ^{+1.7}_{-1.8}$ & $7.5^{+2.2}_{-1.8}\ ^{+0.9}_{-1.4}$\\\hline
		 $\Lambda_{b}(5912)$ & $1/2^{-}$ & $5912.18\pm0.13\pm0.17$ & $<0.66$\\
		 $\Lambda_{b}(5920)$ & $3/2^{-}$ & $5919.90\pm0.19$ & $<0.63$\\\hline
		\end{tabular}
		\label{exp. table}
	\end{center}
\end{table}

We determine the values of the coupling constants $g_{2}^{Q}$ ($Q=c, b$) from the mass differences $\Delta M_{Q}$ of chiral partners in the following way:
First, we separately evaluate the mass differences of chiral partners with spin-$1/2$ and spin-$3/2$ 
as
\begin{align}
	\Delta M_{Q}^{(1/2,\ \mathrm{exp})}\notag&=M_{\Lambda_{Q1}}-M_{\Sigma_{Q}},\\
	\Delta M_{Q}^{(3/2,\ \mathrm{exp})}&=M_{\Lambda_{Q1}^{*}}-M_{\Sigma_{Q}^{*}},\quad (Q=c, b),\label{dM}
\end{align}
where $M_{\Lambda_{Q1}^{(*)}}$ and $M_{\Sigma_{Q}^{(*)}}$ are given by taking the isospin average of relevant masses. Using the values listed in Table~\ref{exp. table}, we obtain
\begin{align}
	\frac{\Delta M_{c}^{(1/2,\ \mathrm{exp})}}{f_{\pi}}&=1.19,\\
	\frac{\Delta M_{c}^{(3/2,\ \mathrm{exp})}}{f_{\pi}}&=1.53
\end{align}
for charm sector.
By taking the spin average of these values, we determine the center value of $g_{2}^{c}$ as
\begin{equation}
	g_{2}^{c}=\frac{1}{3}\left(\frac{\Delta M_{c}^{(1/2,\ \mathrm{exp})}}{f_{\pi}}+2\frac{\Delta M_{c}^{(3/2,\ \mathrm{exp})}}{f_{\pi}}\right)=1.30 \ .
\end{equation}
By taking the violation of heavy-spin symmetry, we evaluate the error as
\begin{equation}
	g_{2}^{c}=1.30_{-|1.19-1.30|}^{+|1.53-1.30|} = 1.30^{+0.23}_{-0.11} \ .
\end{equation}
Similarly, $g_2^b$ is evaluated as
\begin{equation}
g_{2}^{b}=0.980^{+0.090}_{-0.046}\ .
\end{equation}

Let us determine the value of the coupling constant $g_{3}$ from $\Sigma_c \to \Lambda_c \pi$ decays.
We use the experimental values of the full widths of $\Sigma_c^{++}(2455;1/2^+)$, $\Sigma_c^{0}(2455;1/2^+)$, $\Sigma_c^{\ast ++}(2520;3/2^+)$ and $\Sigma_c^{\ast 0}(2520;3/2^+)$ with assuming that the one-pion decay is dominant decay mode for each particle.  We  first calculate four values of the effective couplings 
$g_3(\Sigma_{c}^{(\ast)++}\to\Lambda_{c}^{(\ast)+}\pi^{+})$ and $g_{3}(\Sigma_{c}^{(\ast)0}\to\Lambda_{c}^{(\ast)+}\pi^{-})$ from the corresponding decay widths.
Then, taking the iso-spin average for $J^P = 1/2^+$ and $3/2^+$ separately, we obtain 
\begin{align}
	&g_{3}^{(1/2)}=\frac{g_{3}(\Sigma_{c}^{++}\to\Lambda_{c}^{+}\pi^{+})+g_{3}(\Sigma_{c}^{0}\to\Lambda_{c}^{+}\pi^{-})}{2}=0.673\ , \notag\\
	&g_{3}^{(3/2)}=\frac{g_{3}(\Sigma_{c}^{*++}\to\Lambda_{c}^{+}\pi^{+})+g_{3}(\Sigma_{c}^{*0}\to\Lambda_{c}^{+}\pi^{-})}{2}=0.695\ .
\end{align}
The spin average of the above values are calculated as
\begin{equation}
	g_{3}=\frac{1}{3}\left(g_{3}^{1/2}+2g_{3}^{3/2}\right)=0.688\ .
\end{equation}
We include the systematic error of the spin average and the statistical error of the experimental data as 
\begin{equation}
	g_{3}\ ^{+\left|g_{3}^{3/2}-g_{3}\right|+\mathrm{stat.e.}}_{-\left|g_{3}^{1/2}-g_{3}\right|-\mathrm{stat.e.}}\ ,
\end{equation}
to obtain
\begin{equation}
g_3 = 0.688^{+0.013}_{-0.025}\ .
\end{equation}

We summarize the estimated values of $g_{2}^{Q}$ and $g_{3}$ in Table~\ref{tab:g2 g3}.
\begin{table}[h]
	\caption{Estimated values of the coupling constants $g_2^Q$ and $g_3$.}
	\begin{center}
		\begin{tabular}{ccc} \hline \hline
		 parameter & value\\\hline
		 $g_{2}^{c}$ & $1.30^{+0.23}_{-0.11}$\\
		 $g_{2}^{b}$ & 
		 $0.980^{+0.090}_{-0.046}$\\
		 $g_{3}$ & $0.688^{+0.013}_{-0.025}$\\ \hline
		\end{tabular}
	\end{center}
\label{tab:g2 g3}
\end{table} 
Using the estimated value of $g_3$, we calculate the decay widths of $\Sigma_Q^{(\ast)} \to \Lambda_Q \pi$ as shown in Table~\ref{decay of SigmaC}.
\begin{table}[h]
	\caption{Decay widths $\Sigma_{Q}^{(*)}\to\Lambda_{Q}\pi$ calculated in our model. We use $\Sigma_{c}(2455)^{++}\to\Lambda_{c}^{+}\pi^{+},\ \Sigma_{c}(2455)^{++}\to\Lambda_{c}^{+}\pi^{+},\ \Sigma_{c}(2455)^{++}\to\Lambda_{c}^{+}\pi^{+}$ and $\Sigma_{c}(2455)^{++}\to\Lambda_{c}^{+}\pi^{+}$  to determine the coupling constant $g_{3}$ as explained in the text.}
	\begin{center}
		\begin{tabular}{cccccc} \hline \hline
		 decay modes & our model [MeV] & expt. [MeV] \\\hline
		 $\Sigma_{c}^{++}\to\Lambda_{c}^{+}\pi^{+}$ & $1.96^{+0.07}_{-0.14}$ & $1.89^{+0.09}_{-0.18}$\\
		 $\Sigma_{c}^{+}\to\Lambda_{c}^{+}\pi^{0}$ & $2.28^{+0.09}_{-0.17}$ & $<4.6$\\
		 $\Sigma_{c}^{0}\to\Lambda_{c}^{+}\pi^{-}$ & $1.94^{+0.07}_{-0.14}$ & $1.83^{+0.11}_{-0.19}$\\\hline
		 $\Sigma_{c}^{*++}\to\Lambda_{c}^{+}\pi^{+}$ & $14.7^{+0.6}_{-1.1}$ & $14.78^{+0.30}_{-0.40}$\\
		 $\Sigma_{c}^{*+}\to\Lambda_{c}^{+}\pi^{0}$ & $15.3^{+0.6}_{-1.1}$ & $<17$\\
		 $\Sigma_{c}^{*0}\to\Lambda_{c}^{-}\pi^{0}$ & $14.7^{+0.6}_{-1.1}$ & $15.3^{+0.4}_{-0.5}$\\\hline
		 $\Sigma_{b}^{+}\to\Lambda_{b}^{0}\pi^{+}$ & $6.14^{+0.23}_{-0.45}$ & $9.7^{+3.8}_{-2.8}\ ^{+1.2}_{-1.1}$\\
		 $\Sigma_{b}^{0}\to\Lambda_{b}^{0}\pi^{0}$ & $7.27^{+0.27}_{-0.53}$ & -\\
		 $\Sigma_{b}^{-}\to\Lambda_{b}^{0}\pi^{-}$ & $7.02^{+0.27}_{-0.51}$ & $4.9^{+3.1}_{-2.1}\pm1.1$\\\hline
		 $\Sigma_{b}^{*+}\to\Lambda_{b}^{0}\pi^{+}$ & $11.0^{+0.4}_{-0.8}$ & $11.5^{+2.7}_{-2.2}\ ^{+1.0}_{-1.5}$\\
		 $\Sigma_{b}^{*0}\to\Lambda_{b}^{0}\pi^{0}$ & $12.3^{+0.5}_{-0.9}$ & -\\
		 $\Sigma_{b}^{*-}\to\Lambda_{b}^{0}\pi^{-}$ & $11.9^{+0.4}_{-0.9}$ & $7.5^{+2.2}_{-1.8}\ ^{+0.9}_{-1.4}$\\\hline
		\end{tabular}
		\label{decay of SigmaC}
	\end{center}
\end{table} 
This shows that the obtained widths of $\Sigma_{c}(J^P=1/2^+)$ and $\Sigma_{c}^{*}(3/2^+)$ are consistent with each other even though we used common coupling constant $g_{3}$.
This implies that heavy quark spin violation between them is small.
Furthermore, the predicted widths of $\Sigma_{b}$ and $\Sigma_b^{\ast}$ obtained by the common $g_{3}$ coupling for charm and bottom sectors are consistent with experiments. 
This indicates that the violation of the heavy-quark flavor symmetry is small at this moment, but precise determination of them by future experiments might require the inclusion of heavy-quark flavor violation.
We should note that the predicted widths of $\Sigma_c^{(\ast)+} \to \Lambda_c^+ \pi^0$ are larger than those of their iso-spin partners since the phase space is larger due to the smallness of the mass of $\pi^0$.

\vspace{0.5cm}

\section{$\Lambda_{Q1}^{(\ast)}\to\Lambda_{Q}\pi\pi$ decays}
\label{sec:2pi}

\begin{figure}[h]
\begin{center}
\includegraphics[width=8cm]{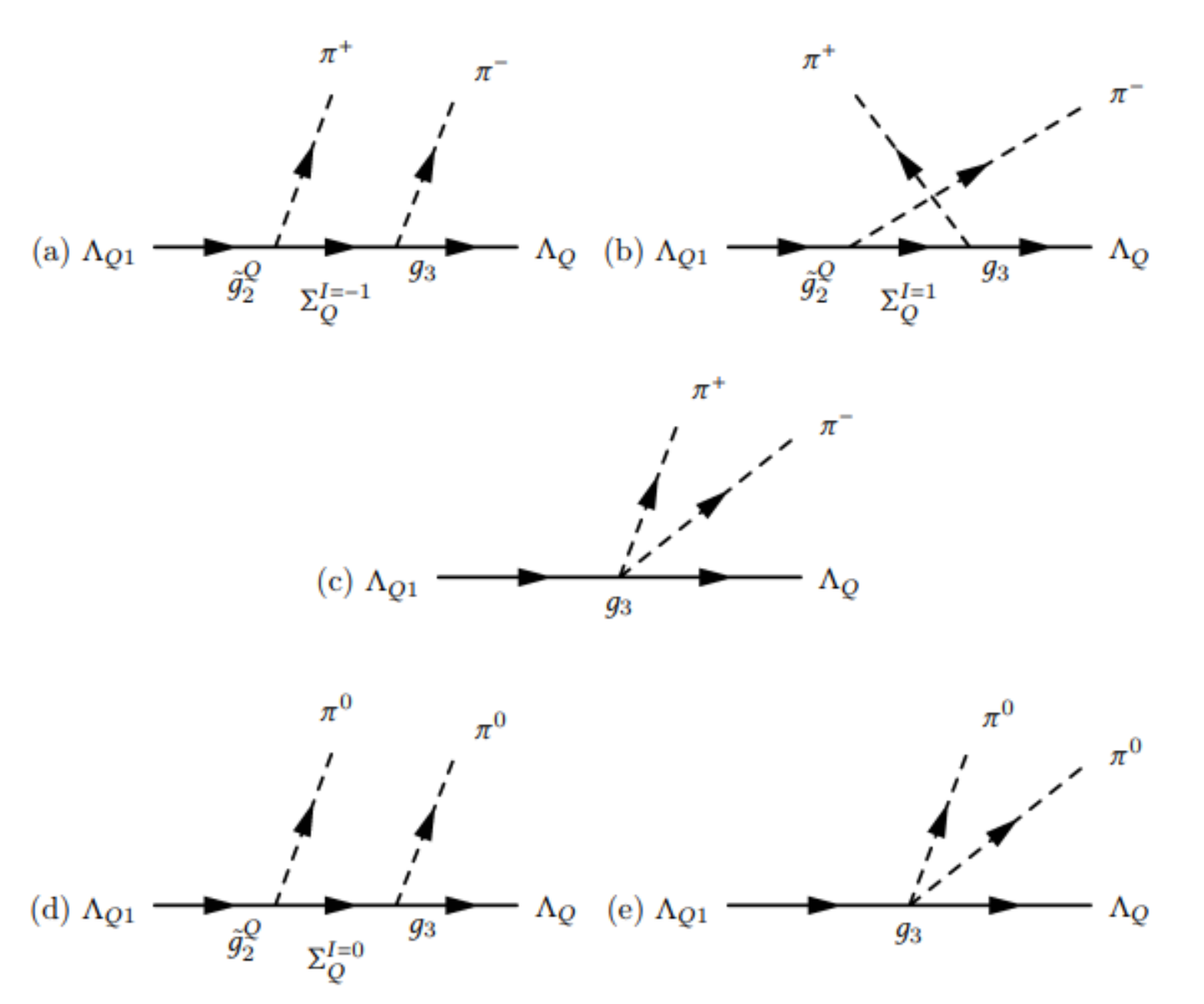}
\end{center}
\caption{Feynman diagrams contributing to $\Lambda_{Q1}\to\Lambda_{Q}\pi\pi$ decay. The effective coupling $\tilde{g}_{2}^Q$ 
is defined as $\tilde{g}_{2}^Q=g_{2}^Q+(h_{1}^{I}+ih_{2})\frac{E_{\pi}}{f_{\pi}}$, 
where $E_{\pi}$ is the relevant pion energy.
Similar Feynman diagrams contribute to $\Lambda_{Q1}^\ast \to \Lambda_Q \pi \pi$ decay. }
\label{fig:FMD}
\end{figure}

In this section, we consider $\Lambda_{Q1}^{(\ast)}\to\Lambda_{Q}\pi\pi$ decays. In Fig.~\ref{fig:FMD}, we plot the relevant diagrams for $\Lambda_{Q1}^{(\ast)}\to\Lambda_{Q}\pi^{+}\pi^{-}$ in our model. 
In the diagrams (a), (b) and (d), $\Sigma_{Q}^{(\ast)}$s appear as intermediate states, while in the diagram (c) and (e), $\Lambda_{Q1}^{(\ast)}$ and $\Lambda_{Q}$ couple to two pions directly. It should be noticed that, due to the chiral partner structure, the coupling constant in (c) and (e) is equivalent to the $\Sigma_{Q}^{(\ast)}\to\Lambda_{Q}\pi$ coupling in (a), (b) and (d). Then, it is not  suitable to drop the contributions in (c) and (e). Actually, as we will show below, They are not neglible for $\Lambda_{c1}^{*}$ and $\Lambda_{b1}^{(*)}$ decays.

\begin{widetext}
From the diagrams in Fig.~\ref{fig:FMD}, the amplitue of $\Lambda_{Q1}\to\Lambda_{Q}\pi^{+}\pi^{-}$ decays is calculated as
\begin{align}
	\mathcal{M}\notag=&-\frac{g_{3}}{\sqrt{3}f_{\pi}^{2}}(p_{2}^{\mu}+p_{3}^{\mu})\bar{u}(p_{1}, t)\left(\gamma_{\mu}+\frac{P_{\mu}}{M}\right)\gamma_{5}u_{1}(P, s)\\\notag
	&-\frac{g_{3}}{\sqrt{3}f_{\pi}}\left\{g_{2}+(h_{1}^{I}+ih_{2})\frac{E_{2}(p_{2})}{f_{\pi}}\right\}S_{f}^{++}(q)\ p_{3}^{\mu}\bar{u}(p_{1}, t)\left(\gamma_{\mu}+\frac{q_{\mu}}{m^{++}}\right)\gamma_{5}(m^{++}+\Slash{q})u_{1}(P, s)\\
	&-\frac{g_{3}}{\sqrt{3}f_{\pi}}\left\{g_{2}+(h_{1}^{I}+ih_{2})\frac{E_{3}(p_{3})}{f_{\pi}}\right\}S_{f}^{0}(k)\ p_{2}^{\mu}\bar{u}(p_{1}, t)\left(\gamma_{\mu}+\frac{k_{\mu}}{m^{0}}\right)\gamma_{5}(m^{0}+\Slash{k})u_{1}(P, s)\ ,
\end{align}
where $P$ is the initial momentum of $\Lambda_{Q1}$, $p_{1}$ the momentum of $\Lambda_{Q}$, $p_{2}$ and $p_{3}$ are the momenta of pions, and $k$ and $q$ the momenta of intermediate $\Sigma_{Q}$s.
$S_{f}$ is the propagator for the intermediate $\Sigma_{Q}$s given by
\begin{equation}
	S_{f}(k)\equiv\frac{1}{m_{\Sigma_{Q}}^{2}-k^{2}+im_{\Sigma_{Q}}\Gamma_{\Sigma_{Q}}},
\end{equation}
where $m_{\Sigma_{Q}}$ and $\Gamma_{\Sigma_{Q}}$ are the
mass and decay width of intermediate $\Sigma_{Q}$. We used isospin-averaged values of masses and decay widths in the present analysis.
Similarly, the amplitude $\Lambda_{Q1}\to\Lambda_{Q}\pi^{0}\pi^{0}$ decays is
\begin{align}
	\mathcal{M}=&-\frac{g_{3}}{\sqrt{3}f_{\pi}^{2}}(p_{2}^{\mu}+p_{3}^{\mu})\bar{u}(p_{1}, t)\left(\gamma_{\mu}+\frac{P_{\mu}}{M}\right)\gamma_{5}u_{1}(P, s)\notag \\
	&-\frac{g_{3}}{\sqrt{3}f_{\pi}}\left\{g_{2}+(h_{1}^{I}+ih_{2})\frac{E_{3}(p_{3})}{f_{\pi}}\right\}S_{f}^{+}(k)\ p_{2}^{\mu}\bar{u}(p_{1}, t)\left(\gamma_{\mu}+\frac{k_{\mu}}{m^{+}}\right)\gamma_{5}(m^{+}+\Slash{k})u_{1}(P, s) \ .
\end{align}
\end{widetext}

\begin{figure}[h]
	\centering
	\includegraphics[width=0.5\textwidth]{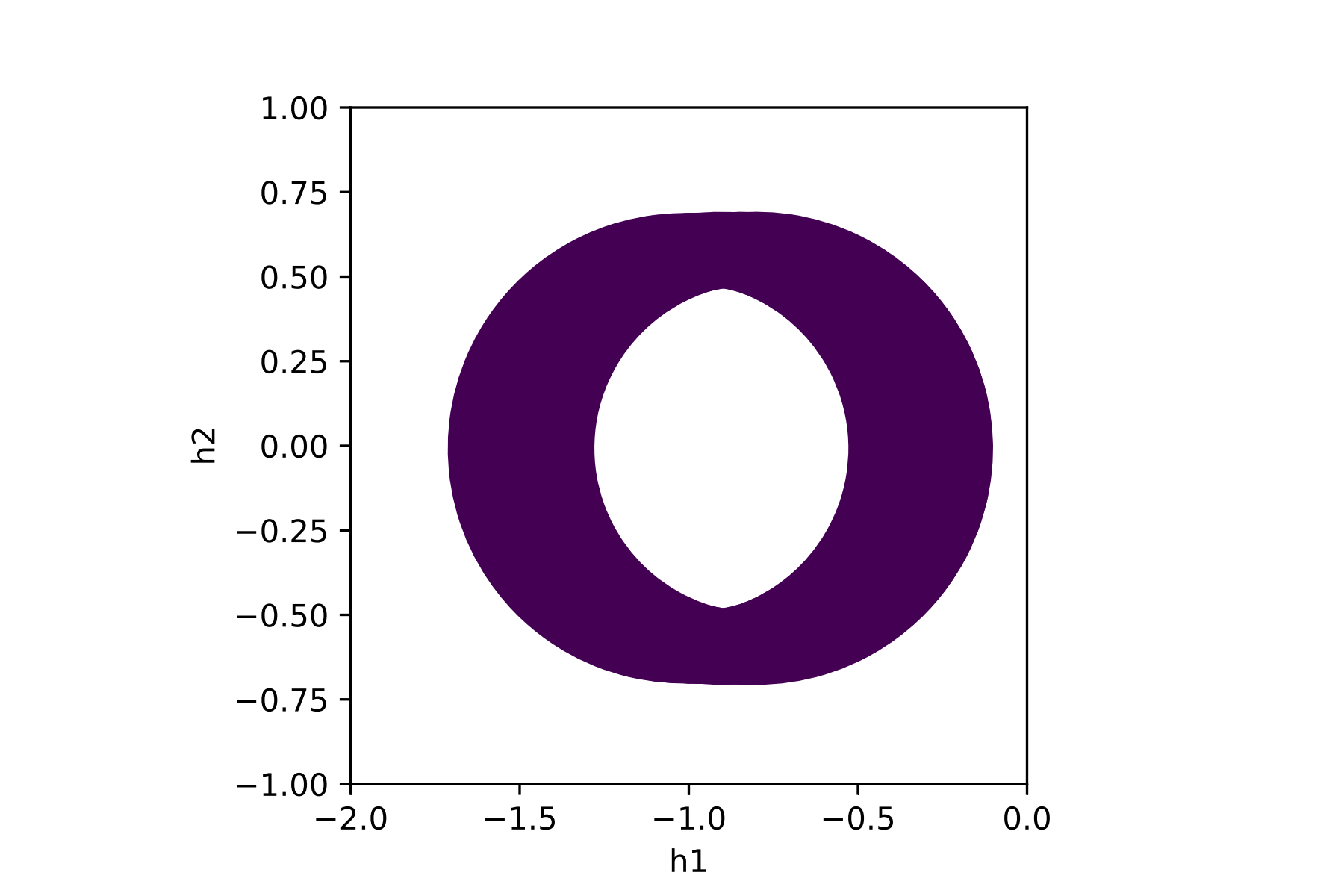}
\caption{Allowed range of $h_{1}^{I}$ and $h_{2}$ shown by purple area.
}
	\label{fig:h1h2}
\end{figure}

We determine the relation between the values of $h_{1}^{I}$ and $h_{2}$ from the full width of $\Lambda_{c1}$ ($\Lambda_{c}(2595)$). Taking into account the errors of $g_{2}^c$, $g_{3}$ and the total width with $\Lambda_{c1}$, we determine
the allowed range of $h_{1}^{I}$ and $h_{2}$ as shown in Fig.~\ref{fig:h1h2}.
Using these values we calculate the two-pion decay widths of $\Lambda_{c}(2625)$, $\Lambda_b(5912)$ and $\Lambda_b(5920)$, which are summarized in Table~\ref{3bd}.
\begin{table}[H]
\caption{Predicted widths of $\Lambda_{Q1}\to\Lambda_{Q}\pi\pi$ decays.
}\label{3bd}
\begin{center}
\begin{tabular}{cccccc} \hline \hline
initial & mode & Our model & expt.\\ 
 & & [MeV] & [MeV] \\
\hline
$\Lambda_{c}(2595)$ & $\Lambda_c\pi^+\pi^-$ & $0.562$-$1.09$  &  \\
 & $\Lambda_c\pi^0\pi^0$ & $1.23$-$2.31$ & \\
 & sum & $1.82$-$3.36$ (input) & $2.59\pm0.30\pm0.47$ \\
\hline
$\Lambda_{c}(2625)$ & $\Lambda_c\pi^+\pi^-$ & $0.0618$-$0.507$ & \\
& $\Lambda_c\pi^0\pi^0$ & $0.0431$-$0.226$ & \\
& sum & $0.106$-$0.733$ & $<0.97$ \\
\hline
$\Lambda_{b}(5912)$ & $\Lambda_b\pi^+\pi^-$ & $(0.67$-$4.4)\times10^{-3}$ &  \\
 & $\Lambda_b\pi^0\pi^0$ & $(1.4$-$6.0)\times10^{-3}$ &  \\
 & sum & $(2.1$-$10)\times10^{-3}$  & $<0.66$ \\
\hline
$\Lambda_{b}(5920)$ & $\Lambda_b\pi^+\pi^-$ & $(0.75$-$13)\times10^{-3}$ & \\
 & $\Lambda_b\pi^0\pi^0$ & $(2.2$-$12)\times10^{-3}$  & \\
 & sum & $(3.0$-$25)\times10^{-3}$ & $<0.63$ \\
\hline
		\end{tabular}		
\end{center}
\end{table} 
The predicted decay width of $\Lambda_{c}(2625)$ is consistent with predictions of a quark model in Ref.~\cite{Nagahiro:2016nsx,Arifi:2017sac}.
We note that 
the predicted decay widths of $\Lambda_b(5912)$ and $\Lambda_b(5920)$ are extremely tiny due to the phase space suppression. 
As we will show in the next section, 
the radiative decay widths for $\Lambda_b(5912)$ and $\Lambda_b(5920)$ are comparable with or even larger than the hadronic decay widths. 

Here we pick up several typical choices of $h_1^I$ and $h_2$, and study the contributions of the diagrams in Fig.~\ref{fig:FMD} and their interferences.
In Table~\ref{typical values}, we list four typical sets of $h_1^I$ and $h_2$ together with the values of $g_2^Q$ and $g_3$.
\begin{table}[h]
	\caption{Four typical parameter sets determined from $\Lambda_{c1}\to\Lambda_c\pi\pi$ decay width.}\label{typical values}
	\begin{center}
		\begin{tabular}{c|ccccc} \hline \hline
		set & $g_{2}^{c}$ & $g_{2}^{b}$ & $g_{3}$ & $h_{1}^{I}$ & $h_{2}$\\\hline
		set 1 & 1.30 & 0.980 & 0.688 & -0.277 & 0\\
		set 2 & 1.30 & 0.980 & 0.688 & -1.45 & 0\\
		set 3 & 1.30 & 0.980 & 0.688 & -0.450 & 0.500\\
		set 4 & 1.30 & 0.980 & 0.688 & -1.00 & -0.500\\\hline
		\end{tabular}
	\end{center}
\end{table} 

\begin{widetext}
Using four sets of parameters in Table~\ref{typical values}, we study contributions of intermediate states to $\Lambda_{Q1}^{(\ast)} \to \Lambda_Q \pi\pi$ decays, which are shown in Tables~\ref{tab:2pi1}-\ref{tab:2pi4}.
\begin{table*}[htbp]
\caption{Contributions of intermediate states to $\Lambda_{c}(2595;1/2^-)\to\Lambda_{c}\pi\pi$ decay.  ``NR(c)'' and ``NR(e)'' in the column for ``intermediate states'' indicate the non-resonant contributions expressed in Figs.~\ref{fig:FMD}(c) and \ref{fig:FMD}(e), respectively.
``$\Sigma_{c}^{0}$(a)'', ``$\Sigma_{c}^{++}$(b)'' and ``$\Sigma_{c}^{+}$(d)'' indicate the resonant contributions in Figs.~\ref{fig:FMD}(a), \ref{fig:FMD}(b) and \ref{fig:FMD}(d), respectively.
``$\Sigma_{c}^{++}$(b) \& $\Sigma_{c}^{0}$(a)'', and so on indicate the contributions of the interferences.
}\label{tab:2pi1}
	\begin{center}
		\begin{tabular}{cccccc} \hline \hline
		 decay mode & intermediate states & set 1 [keV] & set 2 [keV] & set 3 [keV] & set 4 [keV]\\ \hline
		 $\Lambda_{c}(2595;1/2^-)\to\Lambda_{c}^{+}\pi^{+}\pi^{-}$ & NR(c) & 4.10 & 4.10 & 4.10 & 4.10\\
		 & $\Sigma_{c}^{++}$(b) & 344 & 408 & 438 & 302\\
		 & $\Sigma_{c}^{0}$(a) & 390 & 466 & 497 & 344\\
		 & $\Sigma_{c}^{++}$(b) \& $\Sigma_{c}^{0}$(a) & 15.7 & 26.4 & 21.5 & 18.2\\
		 & NR(c) \& $\Sigma_{c}^{++}$(b) & 42.7 & -49.1 & 36.3 & -21.1\\
		 & NR(c) \& $\Sigma_{c}^{0}$(a) & 44.3 & -51.5 & 38.1 & -22.6\\\hline
		 $\Lambda_{c}(2595;1/2^-)\to\Lambda_{c}^{+}\pi^{0}\pi^{0}$ & NR(e) & 4.85 & 4.85 & 4.85 & 4.85\\
		 & $\Sigma_{c}^{+}$(d) & $1.71\times10^{3}$ & $1.82\times10^{3}$ & $2.12\times10^{3}$ & $1.39\times10^{3}$\\
		 & NR(e) \& $\Sigma_{c}^{+}$(d) & 33.1 & -41.6 & 63.7 & -54.6\\\hline
		 total & & $2.59\times10^{3}$ & $2.59\times10^{3}$ & $3.23\times10^{3}$ & $1.97\times10^{3}$\\ \hline
		\end{tabular}
	\end{center}
\end{table*} 
Table~\ref{tab:2pi1} shows the contributions of intermediate states to $\Lambda_{c}(2595;1/2^-)\to\Lambda_{c}\pi\pi$ decay.   More than half of this decay width is provided by the contribution in which $\Sigma_{c}^{+}$ exists as an intermediate state as in Fig.~\ref{fig:FMD}(d). This is because the threshold for $\Lambda_{c}(2595) \to \Sigma_c^+(2455) \pi^0$ decay is open.
Accordingly, non-resonant (NR) contributions in Figs.~\ref{fig:FMD}(c) and (e) are very small.
\begin{table*}[htbp]
\caption{Contributions of intermediate states to $\Lambda_{c}(2625;3/2^-)\to\Lambda_{c}\pi\pi$ decay.  ``NR(c)'' and ``NR(e)'' in the column for ``intermediate states'' indicate the non-resonant contributions expressed in Figs.~\ref{fig:FMD}(c) and \ref{fig:FMD}(e), respectively.
``$\Sigma_{c}^{\ast0}$(a)'', ``$\Sigma_{c}^{\ast++}$(b)'' and ``$\Sigma_{c}^{\ast+}$(d)'' indicate the resonant contributions in Figs.~\ref{fig:FMD}(a), \ref{fig:FMD}(b) and \ref{fig:FMD}(d), respectively.
``$\Sigma_{c}^{\ast++}$(b) \& $\Sigma_{c}^{\ast0}$(a)'', and so on indicate the contributions of the interferences.}
	\begin{center}
		\begin{tabular}{cccccc} \hline \hline
		 decay mode & intermediate states & set 1 [keV] & set 2 [keV] & set 3 [keV] & set 4 [keV]\\ \hline
		 $\Lambda_{c}(2625;3/2^-)\to\Lambda_{c}^{+}\pi^{+}\pi^{-}$ & NR(c) & 58.4 & 58.4 & 58.4 & 58.4 \\
		 & $\Sigma_{c}^{*++}$(b) & 78.2 & 149 & 113 & 97.8\\
		 & $\Sigma_{c}^{*0}$(a) & 76.7 & 157 & 114 & 102\\
		 & $\Sigma_{c}^{*++}$(b) \& $\Sigma_{c}^{*0}$(a) & 7.44 & 22.0 & 12.5 & 13.6\\
		 & NR(c) \& $\Sigma_{c}^{*++}$(b) & 93.2 & -135 & 74.8 & -62.8\\
		 & NR(c) \& $\Sigma_{c}^{*0}$(a) & 92.3 & -138 & 74.3 & -65.3\\\hline
		 $\Lambda_{c}(2625;3/2^-)\to\Lambda_{c}^{+}\pi^{0}\pi^{0}$ & NR(e) & 43.0 & 43.0 & 43.0 & 43.0\\
		 & $\Sigma_{c}^{*+}$(d) & 69.1 & 125 & 97.8 & 83.1\\
		 & NR(e) \& $\Sigma_{c}^{*+}$(d) & 73.2 & -105 & 60.4 & -50.1\\\hline
		 total & & 591 & 177 & 648 & 219\\ \hline
		\end{tabular}
	\end{center}
\end{table*} 
On the other hand, NR contributions to $\Lambda_c(2625)$, $\Lambda_b(5912)$ and $\Lambda_b(5920)$ are comparable to resonant contributions, partly because the threshold for $\Sigma_Q \pi$ decays are not open.
As we stressed, the coupling constant $g_3$ in Figs.~\ref{fig:FMD}(c) and (e) is fixed from $\Sigma_c \to \Lambda_c \pi$ decay based on the chiral partner structure.  Then, the experimental check of non-resonant contributions will give a clue to the chiral symmetry structure.
\begin{table*}[htbp]
\caption{Contributions of intermediate states to $\Lambda_{b}(5912;1/2^-)\to\Lambda_{b}\pi\pi$ decay.  ``NR(c)'' and ``NR(e)'' in the column for ``intermediate states'' indicate the non-resonant contributions expressed in Figs.~\ref{fig:FMD}(c) and \ref{fig:FMD}(e), respectively.
``$\Sigma_{b}^{-}$(a)'', ``$\Sigma_{b}^{+}$(b)'' and ``$\Sigma_{b}^{0}$(d)'' indicate the resonant contributions in Figs.~\ref{fig:FMD}(a), \ref{fig:FMD}(b) and \ref{fig:FMD}(d), respectively.
``$\Sigma_{b}^{+}$(b) \& $\Sigma_{b}^{-}$(a)'', and so on indicate the contributions of the interferences.}
	\begin{center}
		\begin{tabular}{cccccc} \hline \hline
		 decay mode & intermediate states & set 1 [keV] & set 2 [keV] & set 3 [keV] & set 4 [keV]\\ \hline
		 $\Lambda_{b}(5912;1/2^-)\to\Lambda_{b}^{0}\pi^{+}\pi^{-}$ & NR(c) & 0.61 & 0.61 & 0.61 & 0.61 \\
		 & $\Sigma_{b}^{+}$(b) & 0.42 & 2.3 & 0.95 & 1.3\\
		 & $\Sigma_{b}^{-}$(a) & 0.35 & 1.9 & 0.80 & 1.1\\
		 & $\Sigma_{b}^{+}$(b) \& $\Sigma_{b}^{-}$(a) & 0.018 & 0.11 & 0.043 & 0.062\\
		 & NR(c) \& $\Sigma_{b}^{+}$(b) & 0.71 & -1.7 & 0.47 & -0.87\\
		 & NR(c) \& $\Sigma_{b}^{-}$(a) & 0.65 & -1.5 & 0.38 & -0.75\\\hline
		 $\Lambda_{b}(5912;1/2^-)\to\Lambda_{b}^{0}\pi^{0}\pi^{0}$ & NR(e) & 1.40 & 1.40 & 1.40 & 1.40\\
		 & $\Sigma_{b}^{0}$(d) & 0.97 & 5.0 & 2.1 & 2.9\\
		 & NR(e) \& $\Sigma_{b}^{0}$(d) & 1.6 & -3.8 & 1.0 & -1.9\\\hline
		 total & & 6.8 & 4.3 & 7.8 & 3.9\\ \hline
		\end{tabular}
	\end{center}
\end{table*} 
\par
\begin{table*}[htbp]
\caption{Contributions of intermediate states to $\Lambda_{b}(5920;3/2^-)\to\Lambda_{c}\pi\pi$ decay.  ``NR(c)'' and ``NR(e)'' in the column for ``intermediate states'' indicate the non-resonant contributions expressed in Figs.~\ref{fig:FMD}(c) and \ref{fig:FMD}(e), respectively.
``$\Sigma_{b}^{\ast-}$(a)'', ``$\Sigma_{b}^{\ast+}$(b)'' and ``$\Sigma_{b}^{\ast0}$(d)'' indicate the resonant contributions in Figs.~\ref{fig:FMD}(a), \ref{fig:FMD}(b) and \ref{fig:FMD}(d), respectively.
``$\Sigma_{b}^{\ast+}$(b) \& $\Sigma_{b}^{\ast-}$(a)'', and so on indicate the contributions of the interferences.}\label{tab:2pi4}
	\begin{center}
		\begin{tabular}{cccccc} \hline \hline
		 decay mode & intermediate states & set 1 [keV] & set 2 [keV] & set 3 [keV] & set 4 [keV]\\ \hline
		 $\Lambda_{b}(5920;3/2^-)\to\Lambda_{b}^{0}\pi^{+}\pi^{-}$ & NR(c) & 2.4 & 2.4 & 2.4 & 2.4\\
		 & $\Sigma_{b}^{*+}$(b) & 0.87 & 5.2 & 2.1 & 3.0\\
		 & $\Sigma_{b}^{*-}$(a) & 0.80 & 4.8 & 1.9 & 2.8\\
		 & $\Sigma_{b}^{*+}$(b) \& $\Sigma_{b}^{*-}$(a) & 0.040 & 0.27 & 0.10 & 0.15\\
		 & NR(c) \& $\Sigma_{b}^{*+}$(b) & 2.1 & -5.1 & 1.3 & -2.6\\
		 & NR(c) \& $\Sigma_{b}^{*-}$(a) & 2.0 & -4.9 & 1.1 & -2.4\\\hline
		 $\Lambda_{b}(5920;3/2^-)\to\Lambda_{b}^{0}\pi^{0}\pi^{0}$ & NR(e) & 3.5 & 3.5 & 3.5 & 3.5\\
		 & $\Sigma_{b}^{*0}$(d) & 1.3 & 7.3 & 3.0 & 4.2\\
		 & NR(e) \& $\Sigma_{b}^{*0}$(d) & 3.0 & -7.2 & 1.9 & -3.7\\\hline
		 total & & 16 & 6.4 & 17 & 7.2\\ \hline
		\end{tabular}
	\end{center}
\end{table*} 

\end{widetext}

\section{Radiative decays}
\label{sec:rad}

In this section, we consider radiative decays of the heavy baryons. The relevant Lagrangian is given by

~

\begin{align}
	\mathcal{L}_{\mathrm{rad}}\notag&=\frac{r_{1}}{F}{\rm{tr}}\left(\bar{S}_{Q}^{\mu}Q_{\mathrm{light}}S_{Q}^{\nu}+\bar{S}_{Q}^{\mu T}Q_{\mathrm{light}}S_{Q}^{\nu T}\right)F_{\mu\nu}\\\notag
	&+\frac{r_{2}}{F}{\rm{tr}}\left(\bar{S}_{Q}^{\mu}Q_{\mathrm{light}}S_{Q}^{\nu}-\bar{S}_{Q}^{\mu T}Q_{\mathrm{light}}S_{Q}^{\nu T}\right)\tilde{F}_{\mu\nu}\\\notag
	&+\frac{r_{3}}{F^{2}}\bar{\Lambda}_{Q}{\rm{tr}}\left(S_{Q}^{\mu}\tau^{2}MQ_{\mathrm{light}}v^{\nu}-S_{Q}^{\mu T}\tau^{2}M^{\dagger}Q_{\mathrm{light}}v^{\nu}\right)F_{\mu\nu} \notag\\
	& \qquad +{\rm{h.c}} \notag\\
	&+\frac{r_{4}}{F^{2}}\bar{\Lambda}_{Q}{\rm{tr}}\left(S_{Q}^{\mu}\tau^{2}MQ_{\mathrm{light}}v^{\nu}+S_{Q}^{\mu T}\tau^{2}M^{\dagger}Q_{\mathrm{light}}v^{\nu}\right)\tilde{F}_{\mu\nu} \notag\\
& \qquad +{\rm{h.c}},
\label{rad}
\end{align}
where $F_{\mu\nu}$ is the field strength of the photon and $\tilde{F}_{\mu\nu}$ is its dual tensor: $\tilde{F}_{\mu\nu}=(1/2)\epsilon_{\mu\nu\rho\sigma}F^{\rho\sigma}$. $r_{i}$ ($i=1,\ldots,4$) are dimensionless constants, and 
$F$ is a constant with dimension one. In this analysis, we take $F=350$ MeV following Ref.~\cite{Cho:1994vg}.  We note that the values of the constants $r_i$ are of order one based on quark models~\cite{Cho:1994vg}.

Let us first study the electromagnetic intramultiplet transitions governed by the $r_{1}$-term in Eq.~(\ref{rad}). Let $B^{*}$ denotes the decaying baryon with spin-3/2 ($B^{*}=\Lambda_{Q1}^{*},\ \Sigma_{Q}^{*}$), and $B$ the daughter baryon with spin-1/2 ($B=\Lambda_{Q1},\ \Sigma_{Q}$). Then the radiative decay width is given by
\begin{equation}
\Gamma_{B^{*}\to B\gamma} = C_{{B^{*}B\gamma}}^{2}\,
\frac{ 16 \alpha r_1^2 }{9F^2} 
\frac{m_{B}}{m_{B^{*}}}E_{\gamma}^{3}
\end{equation}
where $\alpha$ is the electromagnetic fine structure constant, $E_\gamma$ is the photon energy and $C_{{B^{*}B\gamma}}$ is the Clebsh-Gordon constant given by 
\begin{align}
	C_{\Sigma_{c}^{*++}\Sigma_{c}^{++}\gamma}=C_{\Sigma_{b}^{*+}\Sigma_{b}^{+}\gamma}\notag&=\frac{2}{3},\\\notag
	C_{\Sigma_{c}^{*+}\Sigma_{c}^{+}\gamma}=C_{\Sigma_{b}^{*0}\Sigma_{b}^{0}\gamma}&=\frac{1}{6},\\\notag
	C_{\Sigma_{c}^{*0}\Sigma_{c}^{0}\gamma}=C_{\Sigma_{b}^{*-}\Sigma_{b}^{-}\gamma}&=-\frac{1}{3},\\
	C_{\Lambda_{c1}^{*+}\Lambda_{c1}^{+}\gamma}=C_{\Lambda_{b}^{*0}\Lambda_{b}^{0}\gamma}&=-\frac{1}{6}.
\end{align}
From this, one naively expects that ratios of radiative decay widths are determined from the squares of these constants as
\begin{align}
& C_{\Sigma_{c}^{*++}\Sigma_{c}^{++}\gamma}^{2}:C_{\Sigma_{c}^{*+}\Sigma_{c}^{+}\gamma}^{2}:C_{\Sigma_{c}^{*0}\Sigma_{c}^{0}\gamma}^{2}:C_{\Lambda_{c1}^{*}\Lambda_{c1}\gamma}^{2}\notag\\
& \qquad =16:1:4:1 \ .
\label{r1:ratio}
\end{align}
In Table~\ref{tab:r1}, we show our predictions on the decay widths of $\Lambda_{Q1}^{*}\to\Lambda_{Q1}\gamma$ and $\Sigma_{Q}^{*}\to\Sigma_{Q}\gamma$ comparing with the predictions in Refs.~\cite{Cho:1994vg} and \cite{Jiang:2015xqa}.
The predicted values for $\ \Sigma_{Q}^{*}\to\Sigma_{Q}\gamma$ decay widths are consistent with the ratio in Eq.~(\ref{r1:ratio}), while the values for $\Lambda_{Q1}^{*}\to\Lambda_{Q1}\gamma$ decay widths are much smaller than the ratio in Eq.~(\ref{r1:ratio}).
This is because the mass differences between $\Lambda_{Q1}^\ast$ and $\Lambda_{Q1}$ are quite small generating huge phase space suppression.
We note that our predictions are consistent with the predictions in  Refs.~\cite{Cho:1994vg} and \cite{Jiang:2015xqa}.
\begin{table}[H]
	\begin{center}
\caption{Radiative decay widths of $\Lambda_{Q1}^{*}\to\Lambda_{Q1}\gamma$ and $\Sigma_{Q}^{*}\to\Sigma_{Q}\gamma$ in unit of keV.
The values in the row indicated by ``Predictions" are our predicted values, where $r_1$ is an undetermined parameter of ${\mathcal O}(1)$.
For comparison, we list predictions in Refs.~\cite{Cho:1994vg} and \cite{Jiang:2015xqa}.
}\label{tab:r1}
		\begin{tabular}{cccc}\hline\hline
		decay mode & Predictions & \cite{Cho:1994vg} & \cite{Jiang:2015xqa}\\ 
		& [keV] & [keV] & [keV] \\ 
		\hline
		$\Sigma_{c}^{*++}\to\Sigma_{c}^{++}\gamma$ & $12\ r_{1}^{2}$ & - &  11.6\\
		$\Sigma_{c}^{*+}\to\Sigma_{c}^{+}\gamma$ & $0.75\ r_{1}^{2}$ & - &  0.85  \\
		$\Sigma_{c}^{*0}\to\Sigma_{c}^{0}\gamma$ & $3.1\ r_{1}^{2}$ & - & 2.92\\
		$\Lambda_{c1}^{*+}\to\Lambda_{c1}^{+}\gamma$ & $0.13\ r_{1}^{2}$ & $0.107c_{R}^{2}$ & -\\ \hline
		$\Sigma_{b}^{*+}\to\Sigma_{b}^{+}\gamma$ & $0.42\ r_{1}^{2}$ & - &  0.60\\
		$\Sigma_{b}^{*0}\to\Sigma_{b}^{0}\gamma$ & $0.024\ r_{1}^{2}$ & - &  0.02\\
		$\Sigma_{b}^{*-}\to\Sigma_{b}^{-}\gamma$ & $0.089\ r_{1}^{2}$ & - &  0.06\\
		$\Lambda_{b1}^{*0}\to\Lambda_{b1}^{0}\gamma$ & $0.0013\ r_{1}^{2}$ & - & -\\ \hline
		\end{tabular}
	\end{center}
\end{table}

We next study the $\Lambda_{Q1}^{(*)}\to\Sigma_{Q}^{(*)}\gamma$ decays which concern the $r_2$-term. The decay widths are expressed as
\begin{align}
\Gamma_{\Lambda_{Q1}\to\Sigma_{Q}\gamma} = & \frac{ 16 \alpha r_2^2 }{9 F^2} 
\frac{m_{\Sigma_{Q}}}{m_{\Lambda_{Q1}}}E_{\gamma}^{3} 
\ , \notag\\
\Gamma_{\Lambda_{Q1}\to\Sigma^{*}_Q\gamma} = & \frac{ 8 \alpha r_2^2}{9 F^2}
\frac{m_{\Sigma^{*}_Q}}{m_{\Lambda_{Q1}}}E_{\gamma}^{3} 
\ , \notag\\
\Gamma_{\Lambda_{Q1}^{*}\to\Sigma_Q\gamma} = & \frac{4 r_2^2}{9 F^2}
\frac{m_{\Sigma_Q}}{m_{\Lambda_{Q1}^{*}}}E_{\gamma}^{3}
\ , \notag\\
\Gamma_{\Lambda_{Q1}^{*}\to\Sigma^{*}_Q\gamma} = & \frac{20 \alpha r_2^2}{9 F^2}
\frac{m_{\Sigma^{*}_Q}}{m_{\Lambda_{Q1}^{*}}}E_{\gamma}^{3} \ .
\end{align}
In Table~\ref{tab:r2}, we show our predictions comparing with those in Ref.~\cite{Cho:1994vg}.
\begin{table}[H]
\begin{center}
\caption{Radiative decay widths of $\Lambda_{Q1}^{*}\to\Sigma_{Q}^{(\ast)}\gamma$ in unit of keV.
The values in the row indicated by ``Predictions" are our predicted values, where $r_2$ is an undetermined parameter of ${\mathcal O}(1)$.
For comparison, we list predictions in Ref.~\cite{Cho:1994vg}.
}\label{tab:r2}
		\begin{tabular}{ccc}\hline\hline
		decay mode & Predictions & \cite{Cho:1994vg}\\ 
		& [keV] & [keV] \\ 
\hline
$\Lambda_{c1}^{+}\to\Sigma_{c}^{+}\gamma$ & $250\, r_{2}^2$ & $127\,c_{RS}^{2}$\\
$\Lambda_{c1}^{+}\to\Sigma_{c}^{*+}\gamma$ & $21\, r_{2}^{2}$ & $6\,c_{RS}^{2}$ \\
$\Lambda_{c1}^{*+}\to\Sigma_{c}^{+}\gamma$ & $120\, r_{2}^{2}$ & $58\,c_{RS}^{2}$ \\
$\Lambda_{c1}^{*+}\to\Sigma_{c}^{*+}\gamma$ & $160\, r_{2}^2$ & $54\,c_{RS}^{2}$\\\hline
$\Lambda_{b1}^{0}\to\Sigma_{b}^{0}\gamma$ & $98\, r_{2}^2$ &  -\\
$\Lambda_{b1}^{0}\to\Sigma_{b}^{*0}\gamma$ & $25\, r_{2}^{2}$ & - \\
$\Lambda_{b1}^{*0}\to\Sigma_{b}^{0}\gamma$ & $31\, r_{2}^{2}$ & - \\
$\Lambda_{b1}^{*0}\to\Sigma_{b}^{*0}\gamma$ & $81\, r_{2}^2$ & - \\\hline
\end{tabular}
\end{center}
\end{table}

The $r_3$-term generates the $\Lambda_{Q1}^{(*)}\to\Lambda_{Q}\gamma$ decay, the width of which is expressed as
\begin{equation}
\Gamma_{\Lambda_{Q1}^{(*)}\to\Lambda_Q\gamma}=\frac{8 \alpha r_{3}^{2}f_{\pi}^{2}}{27 F^{4}}\frac{m_{\Lambda_Q}}{m_{\Lambda_{Q1}^{(*)}}}E_{\gamma}^{3}.
\end{equation}
In Table~\ref{tab:r3}, we show our predictions together with the ones in Ref.~\cite{Cho:1994vg}.
\begin{table}[H]
	\begin{center}
\caption{Radiative decay widths of $\Lambda_{Q1}^{*}\to\Lambda_{Q}\gamma$ in unit of keV.
The values in the row indicated by ``Predictions" are our predicted values, where $r_3$ is an undetermined parameter of ${\mathcal O}(1)$.
For comparison, we list predictions in Ref.~\cite{Cho:1994vg}.
}\label{tab:r3}
		\begin{tabular}{ccc}\hline\hline
decay mode & Predictions & \cite{Cho:1994vg}\\ 
 & [keV] & [keV] \\ \hline
		$\Lambda_{c1}\to\Lambda_{c}\gamma$ & $25\, r_{3}^2$ & $191\,c_{RT}^{2}$\\
		$\Lambda_{c1}^{*}\to\Lambda_{c}\gamma$ & $35\, r_{3}^2$ & $253\,c_{RT}^{2}$\\ \hline
		$\Lambda_{b1}\to\Lambda_{b}\gamma$ & $27\, r_{3}^2$ & -\\
		$\Lambda_{b1}^{*}\to\Lambda_{b}\gamma$ & $29\, r_{3}^2$ & -\\ \hline
		\end{tabular}
	\end{center}
\end{table}

The width of $\Sigma_{Q}^{(*)}\to\Lambda_{Q}\gamma$ decay via the $r_4$-term is given by
\begin{equation}
	\Gamma_{\Sigma^{(*)}_Q\to\Lambda_Q\gamma}=\frac{8 \alpha r_{4}^{2}f_{\pi}^{2}}{3 F^{4}}\frac{m_{\Lambda_Q}}{m_{\Sigma^{(*)}_Q}}E_{\gamma}^{3}\ ,
\end{equation}
and the predicted values are shown in Table~\ref{tab:r4}.
\begin{table}[H]
	\begin{center}
\caption{Radiative decay widths of $\Sigma_{Q}^{*}\to\Lambda_{Q}\gamma$ in unit of keV.
The values in the row indicated by ``Predictions" are our predicted values, where $r_4$ is an undetermined parameter of ${\mathcal O}(1)$.
For comparison, we list predictions in Ref.~\cite{Jiang:2015xqa}.
}\label{tab:r4}
		\begin{tabular}{ccc}\hline\hline
decay mode & Predictions & \cite{Jiang:2015xqa} \\ 
 & [keV] & [keV] \\ \hline
		$\Sigma_{c}^{+}\to\Lambda_{c}^{+}\gamma$ & $43\ r_{4}^{2}$ & 164 \\
		$\Sigma_{c}^{*+}\to\Lambda_{c}^{+}\gamma$ & $110\ r_{4}^{2}$ & 893 \\ \hline
		$\Sigma_{b}^{0}\to\Lambda_{b}^{0}\gamma$ & $74\ r_{4}^{2}$ & 288 \\
		$\Sigma_{b}^{*0}\to\Lambda_{b}^{0}\gamma$ & $99\ r_{4}^{2}$ & 435 \\ \hline
		\end{tabular}
	\end{center}
\end{table}

\section{A summary and discussions}
\label{sec:summary}

We constructed an effective hadronic model regarding 
$\Lambda_{Q1} = \left\{ \Lambda_c(2595,J^P=1/2^-) \,,\, \Lambda_b(5912,1/2^-) \right\}$ and $\Lambda_{Q1}^\ast = \left\{ \Lambda_c^\ast(2625,3/2^-) \,,\, \Lambda_b^\ast(5920,3/2^-) \right\}$
as chiral partners to $\Sigma_{Q} = \left\{ \Sigma_c(2455,1/2^+)\,,\,  \Sigma_b(1/2^+) \right\}$ and $\Sigma_{Q}^{\ast} = \left\{ \Sigma_c^\ast(2520,3/2^+)\,,\,\Sigma_b^\ast(3/2^+) \right\}$, respectively, based on the chiral symmetry and heavy-quark spin-flavor symmetry.
We determined the model parameters from the experimental data for relevant masses and 
decay widths of $\Sigma_c(2455,1/2^+)$, $\Sigma_c^{\ast}(2520,3/2^+)$ and $\Lambda_c(2595,1/2^-)$.
Then, we studied the decay widths of $\Lambda_c^\ast(2625)$, $\Lambda_b(5912)$ and $\Lambda_b(5920)$.
We showed that the coupling constant for non-resonant contributions depicted in Figs.~\ref{fig:FMD}(c) and (e) is fixed from the $\Sigma_{c}^{(\ast)}$ decays reflecting the chiral partner structure.
As a result, the decay of $\Lambda_c(2595)$ is dominated by the resonant contribution through $\Sigma_c(2455)$ depicted in Fig.~\ref{fig:FMD}(d), since the threshold of $\Lambda_c(2595) \to \Sigma_c^+(2455) \pi^0$ decay is open.
We found non-resonant contributions depicted in Figs.~\ref{fig:FMD}(c) and (e) 
are comparable to resonant contributions for $\Lambda_c(2625)$, $\Lambda_b(5912)$ and $\Lambda_b(5920)$, 
partly because the threshold for $\Sigma_Q \pi$ decays are not open.
Our result indicates that studying non-resonant contributions will give a clue to understand the chiral partner structure for single heavy baryons.

We also studied the radiative decays of $\Sigma_{Q}^{(\ast)}$ and $\Lambda_{Q1}^{(\ast)}$ using effective interaction Lagrangians in Eq.~(\ref{rad}).
We showed that there is a relation among $\Sigma_Q^\ast \to \Sigma_Q \gamma$ and $\Lambda_{Q1}^\ast \to \Lambda_{Q1} \gamma$ decays reflecting the chiral partner structure, which can be checked in future experiments.

In Table~\ref{tab:summary}, we summarize our predictions of the decay widths of single heavy baryons.
We expect that comparison of these predictions with experimental data will give some clues to understand the chiral structure of single heavy baryons.
We note that, since the hadronic decays of $\Lambda_{b1}(5912)$ and $\Lambda_{b1}^\ast(5920)$ are suppressed by the small phase space factors, radiative decays may be dominant modes.
\renewcommand{\arraystretch}{1.3}
\begin{table}[htbp]
\caption{Predicted decay widths of single heavy baryons (SHBs). The row indicated by ``Our model'' shows the predictions of the present analysis. The row indicated by ``exp.'' shows the experimental values for the full width of the relevant SHBs, in which ``-'' implies no experimental data. 
} \label{tab:summary}
\begin{center}
\begin{tabular}{ccccc} \hline \hline
SHB & $J^{P}$ & decay & Our model & exp. \\
 & & modes  & [MeV] & [MeV] \\
\hline
$\Sigma_{c}^{++}$ & $1/2^{+}$ & $\Lambda_{c}\pi^{+}$ & $1.96^{+0.07}_{-0.14})$ & $1.89^{+0.09}_{-0.18}$\\
\hline
$\Sigma_{c}^{+}$ & $1/2^{+}$ & $\Lambda_{c}\pi^{0}$ & $2.28^{+0.09}_{-0.17}$ & \multirow{2}{*}{$<4.6$}\\
 & & $\Lambda_{c}\gamma$ & $0.043\ r_{4}^{2}$ &\\
\hline
$\Sigma_{c}^{0}$ & $1/2^{+}$ & $\Lambda_{c}\pi^{-}$ & $1.94^{+0.07}_{-0.14}$ & $1.83^{+0.11}_{-0.19}$\\\hline
$\Sigma_{c}^{\ast++}$ & $3/2^{+}$ & $\Lambda_{c}\pi^{+}$ & $14.7^{+0.6}_{-1.1}$ & \multirow{2}{*}{$14.78^{+0.30}_{-0.40}$} \\
 & & $\Sigma_{c}^{++}\gamma$ & $0.012\,r_1^2$ & \\
\hline
$\Sigma_{c}^{\ast+}$ & $3/2^{+}$ & $\Lambda_{c}\pi^{0}$ & $15.3^{+0.6}_{-1.1}$ & \multirow{3}{*}{$<17$} \\
 & & $\Sigma_{c}^{+}\gamma$ & $0.75  \, r_1^2 \times 10^{-3}$ & \\
 & & $\Lambda_{c}\gamma$ & $0.11\ r_{4}^{2}$ &\\
\hline
$\Sigma_{c}^{\ast0}$ & $3/2^{+}$ & $\Lambda_{c}\pi^{-}$ & $14.7^{+0.6}_{-1.1}$ & \multirow{2}{*}{$15.3^{+0.4}_{-0.5}$} \\
 & & $\Sigma_{c}^{0}\gamma$ & $3.1 \, r_1^2 \times 10^{-3}$ & \\
 \hline
$\Lambda_{c1}$ & $1/2^{-}$ & $\Lambda_{c}\pi^+\pi^-$ & $0.562$-$1.09$ & \multirow{5}{*}{\footnotesize $2.59\pm0.30\pm0.47$} \\
 & & $\Lambda_{c}\pi^0\pi^0$ & $1.23$-$2.31$ & \\
 & & $\Sigma_c^+\gamma$ & $0.25\,r_2^2$ & \\
 & & $\Sigma_c^{\ast+}\gamma$ & $0.021\,r_2^2$ & \\
 & & $\Lambda_c\gamma$ & $0.025\,r_3^2$ & \\
\hline
$\Lambda_{c1}^\ast$ & $3/2^{-}$ & $\Lambda_{c}\pi^+\pi^-$ & $0.0618$-$0.507$ & \multirow{6}{*}{$<0.97$} \\
 & & $\Lambda_{c}\pi^0\pi^0$ & $0.0431$-$0.226$ & \\
 & & $\Lambda_{c1}\gamma$ & $0.13 \, r_1^2 \times 10^{-3}$ & \\
 & & $\Sigma_c^+\gamma$ & $0.12\,r_2^2$ & \\
 & & $\Sigma_c^{\ast+}\gamma$ & $0.16\,r_2^2$ & \\
 & & $\Lambda_c\gamma$ & $0.035\,r_3^2$ & \\
\hline
$\Sigma_{b}^{+}$ & $1/2^{+}$ & $\Lambda_{b}\pi^{+}$ & $6.14^{+0.23}_{-0.45}$ & $9.7^{+3.8}_{-2.8}\ ^{+1.2}_{-1.1}$\\
\hline
$\Sigma_{b}^{0}$ & $1/2^{+}$ & $\Lambda_{b}\pi^{0}$ & $7.27^{+0.27}_{-0.53}$ & \multirow{2}{*}{-}\\
 & & $\Lambda_{b}^{0}\gamma$ & $0.074 \, r_4^2$ & \\
\hline
$\Sigma_{b}^{-}$ & $1/2^{+}$ & $\Lambda_{b}\pi^{-}$ & $7.02^{+0.27}_{-0.51}$ & $4.9^{+3.1}_{-2.1}\pm1.1$\\\hline
$\Sigma_{b}^{*+}$ & $3/2^{+}$ & $\Lambda_{b}^{0}\pi^{+}$ & $11.0^{+0.4}_{-0.8}$ & \multirow{2}{*}{$11.5^{+2.7}_{-2.2}\ ^{+1.0}_{-1.5}$} \\
 & & $\Sigma_{b}^{+}\gamma$ & $0.42 \, r_1^2 \times 10^{-3} $ & \\
 \hline
$\Sigma_{b}^{*0}$ & $3/2^{+}$ & $\Lambda_{b}\pi^{0}$ & $12.3^{+0.5}_{-0.9}$ & \multirow{3}{*}{-} \\
 & & $\Sigma_{b}^{+}\gamma$ & $0.024 \, r_1^2 \times 10^{-3} $ & \\
 & & $\Lambda_{b}\gamma$ & $0.074\ r_{4}^{2}$ &\\
\hline
$\Sigma_{b}^{*-}$ & $3/2^{+}$ & $\Lambda_{b}\pi^{-}$ & $11.9^{+0.4}_{-0.9}$ & \multirow{3}{*}{$7.5^{+2.2}_{-1.8}\ ^{+0.9}_{-1.4}$} \\
 & & $\Sigma_{b}^{-}\gamma$ & $0.089 \, r_1^2 \times 10^{-3}$ & \\
 & & $\Lambda_{b}\gamma$ & $0.099\ r_{4}^{2}$ &\\
\hline
$\Lambda_{b1}$ & $1/2^{-}$ & $\Lambda_{b}\pi^+\pi^-$ & $(0.67$-$4.4)\times10^{-3}$  & \multirow{5}{*}{$<0.66$} \\
 & & $\Lambda_{b}\pi^0\pi^0$ & $(1.4$-$6.0)\times10^{-3}$ & \\
 & & $\Sigma_b^0\gamma$ & $0.098\,r_2^2$ & \\
 & & $\Sigma_b^{\ast0}\gamma$ & $0.025\,r_2^2$ & \\
 & & $\Lambda_b\gamma$ & $0.027\,r_3^2$ & \\
\hline
$\Lambda_{b1}^\ast$ & $3/2^{-}$ & $\Lambda_{b}\pi^+\pi^-$ & $(0.75$-$13)\times10^{-3}$   & \multirow{6}{*}{$<0.63$} \\
 & & $\Lambda_{b}\pi^0\pi^0$ & $(2.2$-$12)\times10^{-3}$ & \\
 & & $\Lambda_{b1}\gamma$ & $0.0013 \, r_1^2 \times 10^{-3}$ & \\
 & & $\Sigma_b^0\gamma$ & $0.031\,r_2^2$ & \\
 & & $\Sigma_b^{\ast0}\gamma$ & $0.081\,r_2^2$ & \\
 & & $\Lambda_b\gamma$ & $0.029\,r_3^2$ & \\
\hline
\end{tabular}
\end{center}
\end{table} 
\renewcommand{\arraystretch}{1}

Several comments are in order.

The present model does not include the decay $\Lambda_{Q1}\to\Lambda_{Q}\pi\pi$ via $\Sigma_{Q}^{*}$, which needs 
two pions in the $D$-wave in the heavy quark limit.
We expect that such decays are suppressed compared with the decays having two pions in the $S$-wave.
Similarly, we expect that the decay $\Lambda_{Q1}^{*}\to\Lambda_{Q}\pi\pi$ via $\Sigma_{Q}$ is also suppresed.

It is interesting to extend the present model including only two flavors to the one with the strange quark in addition based on the chiral SU(3)$_L \times$SU(3)$_R$ symmetry.
In the case, the flavor $\bar{3}$ representation including $\Lambda_{Q1}$ becomes the chiral partner to the flavor $6$ representation including $\Sigma_{Q}$. 
We leave the analysis in future publication.

\subsection*{Acknowledgments}

This work was supported partly by JSPS KAKENHI Grant Number JP16K05345.
We would like to thank Daiki Suenaga for useful discussions and comments.

\end{document}